\documentclass[aps,pra,twocolumn,showpacs]{revtex4}
\usepackage{mathrsfs}
\usepackage{epsfig}
\usepackage{graphicx}
\usepackage{amsfonts}
\usepackage[figuresright]{rotating}
\usepackage{amssymb}
\usepackage{amsmath}
\usepackage{dcolumn}
\usepackage{bm}

\def\avg#1{\langle#1\rangle}

\def\be{\begin{equation}} \def\ee{\end{equation}}
\def\bea{\begin{eqnarray}} \def\eea{\end{eqnarray}}

\def\nn{\nonumber}

\begin{document}
\title{Complex and real unconventional Bose-Einstein condensations
in high orbital bands}
\author{Zi Cai}
\affiliation{Department of Physics, University of California, San
Diego, CA92093}
\author{Congjun Wu}
\affiliation{Department of Physics, University of California, San
Diego, CA92093}
\begin{abstract}
We perform the theoretical study on the unconventional Bose-Einstein
condensations (UBEC) in the high bands of optical lattices observed
by Hemmerich's group. These exotic states are characterized by
complex-valued condensate wavefunctions with nodal points, or
real-valued ones with nodal lines, thus are beyond the {\it ``no-node''}
paradigm of the conventional BECs. A quantum phase transition is driven by
the competition between the single particle band and interaction energies. The
complex UBECs spontaneously break time-reversal symmetry, exhibiting
a vortex-antivortex lattice structure.
\end{abstract}
\maketitle

Quantum wavefunctions are generally complex-valued.
However, the usual ground state wavefunctions of bosons are very restricted
because they are positive-definite as stated in the {\it
``no-node''} theorem \cite{feynman1972}. This theorem applies under
very general conditions: the kinetic energy is unfrustrated (e.g.
the Laplacian-type); the single particle potential can be arbitrary;
the two-body interaction depends only on coordinates.
Mathematically, it is a direct consequence of the Perron-Frobenius
theorem of matrix analysis \cite{Bapat1997}. This theorem implies
that time-reversal (TR) symmetry cannot be spontaneously broken in
various ground states of bosons, including superfluid, Mott-insulating,
and supersolid  states.

The ``no-node'' theorem, however, is a ground state property, thus
it does not apply to meta-stable excited states of bosons.
This opens up a possibility for {\it ``unconventional''} states
of bosons beyond the ``no-node'' paradigm \cite{wu2009}.
Similarly to unconventional superconductors, in unconventional
Bose-Einstein condensations (UBEC), the condensate
wavefunctions form non-trivial representations of the lattice
symmetry groups.
However, a major difference exists.
Cooper pairs have the center of mass motion and the relative motion
between two electrons of the pair.
In unconventional superconductors, it is the relative motion
that is non-trivial.
The degree of freedom of the relative motion does not exist in the
single boson BEC.
In UBECs, the condensate wavefunctions are non-trivial.

Considerable efforts have been made to study unconventional states
of bosons both experimentally and theoretically. Among the most
exciting achievements are the realizations of the meta-stable
excited states of bosons in high orbital bands
\cite{sebby-strabley2006,mueller2007,wirth2011,olschlager2011},
which leads to the opportunity to the study  of the UBECs
\cite{liu2006,isacsson2005,alon2005,wu2006,stojanovic2008,kuklov2006,
martikainen2011,Lewenstein2011}, and other exotic properties
\cite{wu2007,scarola2005,xu2007,challis2009,lixp2011,zhou2011}.
Below are some recent experimental results. Sebby-Strabley {\it et
al.} succeeded in pumping a large fraction of bosons into the
excited bands in a double-well lattice \cite{sebby-strabley2006}.
Mueller {\it et al.} observed the quasi-1D phase coherence pattern
by exciting bosons into the $p$-orbital bands in the cubic lattice
\cite{mueller2007}. An important progress was made by the group of
Hemmerich \cite{wirth2011}: the UBECs in the $sp$-hybridized orbital
bands were realized in a checkerboard-like lattice, which allows to
establish the fully cross-dimensional coherence. More recently,
UBECs in even higher orbital bands have been observed in the same
group \cite{olschlager2011}.

In this paper, we present the theoretical study on UBECs observed in
the second, or, the first excited band, of the checkerboard optical
lattice. This band is of a hybridized nature between the $s$-orbital
of the shallower sites and the $p$-orbitals of the deeper sites. The
lattice asymmetry favors a real-valued condensate wavefunction with
nodal lines, while interactions favor a complex-valued one with
nodal points. By solving the Gross-Pitaevskii (GP) equation for
these meta-stable condensates, we find that tuning the lattice
asymmetry drives the phase transition between these two types of
UBECs in a good agreement with experimental observations.

\begin{figure}[htb]
\includegraphics[width=0.9\linewidth]{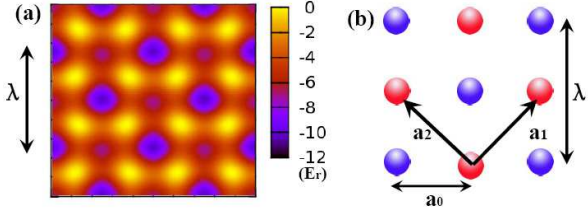} \caption{(a) The
optical lattice with the reflection symmetry with respect to the
$x$-axis and the parameter values: $\eta=0.95$, $\epsilon=0.81$,
$\theta=95.4^\circ$, $\alpha=\alpha_0=36^\circ$ and $V_0=6.2E_r$.
The $A$-sites have deeper potential depth than those of $B$-sites.
(b) The basis vectors of the double-well lattice.}
\label{fig:lattice}
\end{figure}

We introduce the optical lattice employed in the experiment
\cite{wirth2011}. Each unit cell consists of two sites with
different depths (denoted $A$ and $B$ below) as shown in
Fig. \ref{fig:lattice} (a). (A similar lattice potential with
different parameters has been plotted in Ref.\cite{wirth2011}). The
lattice is constructed by the interference pattern of phase coherent
laser beams along $\pm x$ and $\pm y$-directions generated from a
single laser through beam splitters and reflectors. The optical
potential reads \bea V(x,y)&=&-\frac{V_0}4|(\hat{z}\cos\alpha
+\hat{y}\sin\alpha) e^{ik_lx}+\epsilon\hat{z}e^{-ik_lx} \nn \\
&+&\eta e^{i\theta}\hat{z}(e^{ik_ly}+\epsilon e^{-ik_ly})|^2,
\label{eq:optical} \eea where $\hat y$ and $\hat z$ are unit vectors
describing light polarizations; $k_l$ is the laser wavevector;
$\epsilon<1$ and $\eta<1$ describe the imperfect reflection and
transmission efficiencies; $\theta$ is the phase difference between beams along
$x$ and $y$-directions; $\alpha$ is used to tune the
lattice asymmetry by rotating the light polarization out of the
$\hat z$-direction.

The point group symmetry of this lattice is analyzed below. We start
from the ideal case of $\epsilon=1$ with $\alpha=0^\circ$ and
$\theta=90^\circ$, at which $A$ and $B$-sites are equivalent. At
$\eta<1$, the lattice has the reflection symmetries with respect to
both the $x$ and $y$-axes, thus the lattice is orthorhombic. Next we
keep $\epsilon=1$ and $\alpha=0^\circ$ but set $\theta$ away from
$90^\circ$. Then the unit cell includes both $A$ and $B$-sites. The
primitive lattice vectors are $a_0 (\hat e_x \pm \hat e_y)$ where
$a_0=\pi/k_l$ as shown in Fig. \ref{fig:lattice} (b). The optical
potential becomes $V=-\frac{V_0}{2} \big(\cos 2k_l x +\eta^2 \cos
2k_ly + 4 \eta \cos\theta \cos k_l x \cos k_l y \big)$. $\theta$
controls the potential difference between $A$ and $B$ sites. The
point group symmetry remains orthorhombic. Now we move to the
realistic case of $\epsilon<1$. The unit cell remains
double-well-shaped and the primitive lattice vectors are the same.
However, the orthorhombic symmetry is broken and there is no point
group symmetry for general values of parameters. This asymmetry can
be partially compensated by setting $\alpha_0=\cos^{-1} \epsilon$.
We denote this configuration as ``symmetric'' and other ones with
$\alpha\neq \alpha_0$ as ``asymmetric'' below. The symmetric lattice
potential becomes $V=-\frac{V_0}{2}\epsilon \big(\epsilon \cos 2k_l
x+\eta^2\cos 2k_l y \big) - V_0 \eta \epsilon \cos k_l x \big[\cos
(k_l y +\theta) +\epsilon^2 \cos (k_l y -\theta) \big]$, which has
the reflection symmetry with respect to the $x$-axis but not to the
$y$-axis.

Next we calculate the band structures.
The reciprocal lattice vectors are defined as
$\vec G_{m,n}=m \vec b_1 + n\vec b_2$ with
$\vec b_{1,2} = (\pm \frac{\pi}{a}, \frac{\pi}{a})$.
The single particle Hamiltonian reads as $H_0=-\hbar^2 \vec \nabla^2/(2M)+V(r)$
where $M$ is the boson mass.
Using the plane wave basis, the diagonal matrix elements are
$\langle\vec k+\vec G_{mn}|H_0|\vec k+\vec G_{mn}\rangle
=E_r\{[a k_x/\pi +(m-n)]^2 +[a k_y/\pi +(m+n)]^2\}$, where
$E_r=\hbar^2 \pi ^2/(2Ma^2)$ is the recoil energy.
The off-diagonal matrix elements read
\bea
\langle\vec k |H_0|\vec k+\vec G_{\pm1,0}\rangle
&=&-\frac{V_0}{4}\eta\epsilon(\cos\alpha e^{\mp i\theta}+ e^{\pm i\theta}),
\nn \\
\langle\vec k|H_0|\vec k+\vec G_{0,\pm1}\rangle
&=&-\frac{V_0}{4}\eta(\cos\alpha
e^{\pm i\theta}+\epsilon^2e^{\mp i\theta}),\nn\\
\langle \vec k |H_0| \vec k+\vec G_{\pm1,\mp1}\rangle
&=&-\frac{V_0}{4}\epsilon\cos\alpha,\nn \\
\langle \vec k|H_0|\vec k+\vec G_{\pm1,\pm1}\rangle
&=&-\frac{V_0}{4}\epsilon\eta^2\cos\alpha.
\label{eq:energy}
\eea

\begin{figure}[htb]
\includegraphics[width=0.50\linewidth]{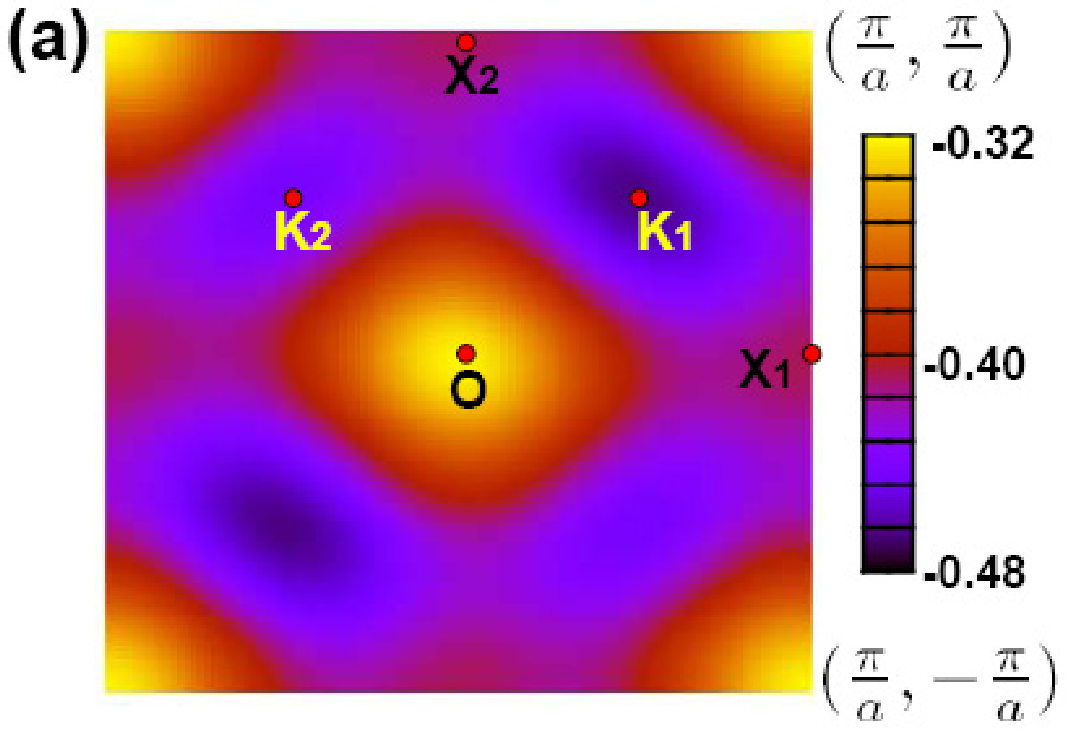}
\includegraphics[width=0.47\linewidth]{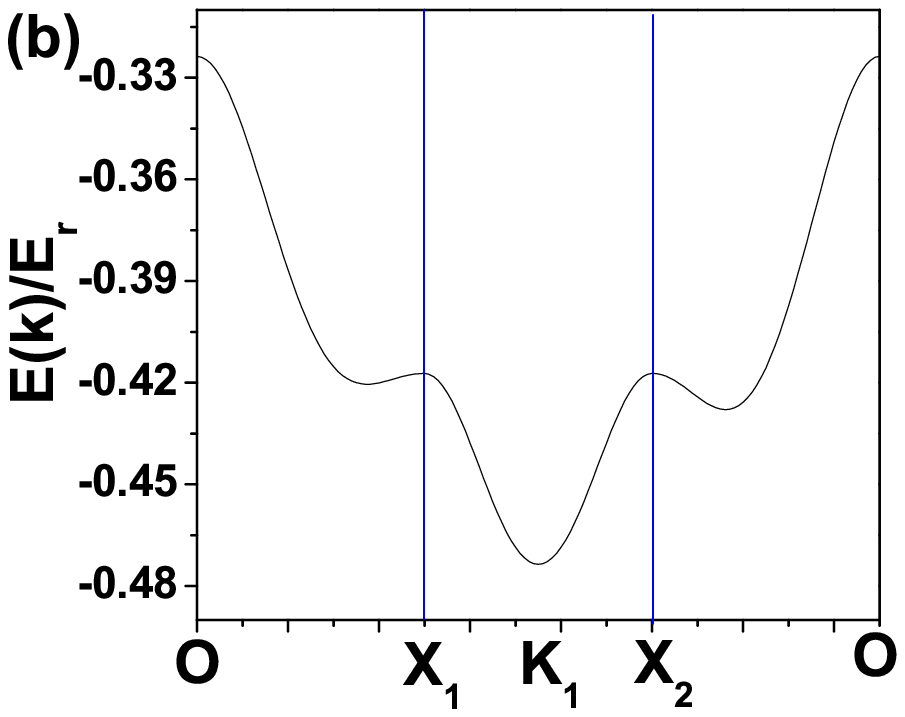}
\caption{(a) The energy spectra for the second band, the parameter
values are the same as Fig.\ref{fig:lattice} except
$\alpha=0^\circ$. (b) The spectra of (a) along the line from
$(0,\frac{\pi}{a})$ to $(\frac{\pi}{a},0)$.} \label{fig:band}
\end{figure}
We focus on the second band into which bosons are pumped
\cite{wirth2011}. There are four points in the Brillouin zone (BZ),
{\it i.e.}, $O=(0,0), ~K_{1,2}=(\pm
\frac{\pi}{2a_0},\frac{\pi}{2a_0})$, and
$X=(\frac{\pi}{a_0},\frac{\pi}{a_0})$, at which the Bloch
wavefunctions are TR invariant, and thus real-valued. The band
spectra are symmetric with respect to these points, which means that
they are local energy extrema or saddle points. For the symmetric
lattice with $\alpha=\alpha_0$, the second band has doubly
degenerate energy minima of the states $\psi_{K_{1,2}}$ located at
$K_{1,2}$, respectively. For the asymmetric case, the degeneracy
between $\psi_{K_{1}}$ and $\psi_{K_{2}}$ are lifted. For
$\alpha<(>)\alpha_0$, $K_{1} (K_2)$ become the band minimum,
respectively. The energy spectra of $\alpha=0$ is shown in  Fig.
\ref{fig:band} (a) and (b) (A similar energy spectrum with different
parameters has been plotted in Ref.\cite{wirth2011}).

The real space distributions of $\psi_{K_{1,2}}$ are also calculated.
Their nodal lines pass the centers of the deeper sites of $A$.
Thus the orbital component on the $A$-sites is of the $p$-type
and that on the shallower sites of $B$ is of the $s$-type.
In fact, the $p$-orbital configurations of $\psi_{K_{1,2}}(\vec r)$
in the $A$-sites are actually not exactly along the directions
of $\hat e_x\pm \hat e_y$ because of the lack of the tetragonal symmetry.
This point is mostly clear in the case of strong  potentials
so that we can define local orbitals on each site.
Even for the symmetric lattice, the $p_x$ and $p_y$-orbitals
on the $A$-sites can be defined according to their parities
under the reflection with respect to the $x$-axis.
However, they are non-degenerate.
The orbital components of $\psi_{K_{1,2}}(\vec r)$
are nearly the same on $A$-sites, {\it i.e.}, mostly the lower
energy $p$-orbital slightly hybridized with the higher one.
The orthogonality of these two states 
comes from their different lattice
momenta.

Interactions determine the configurations of UBECs
in the presence of degenerate band minima.
Any linear superposition among them gives rise to the condensate
wavefunctions with the same kinetic energy.
However, interactions break this degeneracy.
Previous studies on $p$-orbital BECs based on tight-binding models predicted
linear superpositions between two Bloch wavefunctions at degenerate band
minima with a phase difference $\pm\frac{\pi}{2}$.
Such a condensate breaks TR symmetry spontaneously \cite{liu2006,wu2009}.
Bosons on $p$-orbital sites aggregate into
the $p_x\pm ip_y$ states to reduce their repulsive interaction energy.
This is a result of the second Hund's rule: complex $p$-orbitals are
spatially more extended than the real orbitals, and thus bosons have
more room to avoid each other.

\begin{figure}[htb]
\includegraphics[width=0.47\linewidth]{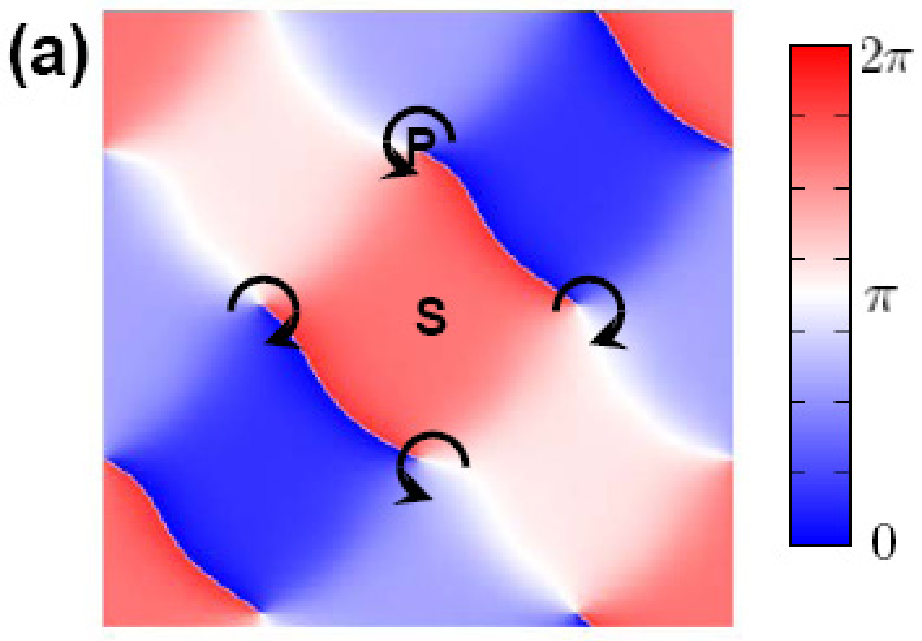}
\includegraphics[width=0.50\linewidth]{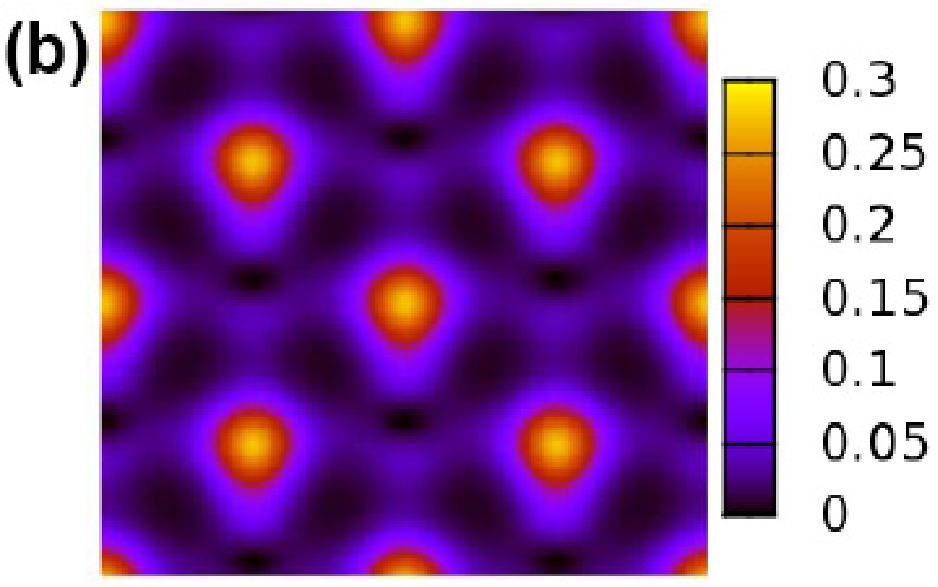}
\caption{The distributions of (a) the phase and (b) density patterns
of the complex UBEC. The parameter values are the same as in
Fig.\ref{fig:lattice} except $g \rho_0=0.6 E_r$ and
$\alpha=36^\circ$. The vortex and anti-vortex cores are located in
the centers of $A$-sites.} \label{fig:phase}
\end{figure}

The optical potential in the current experiment is shallow, thus the
system is in the weak correlation regime \cite{wirth2011}. Instead
of the tight-binding model, we use the GP equation. Because of the
absence of the lattice potential along the $z$-axis, we neglect the
$z$-dependence of the condensate wavefunction. We only consider its
distribution $\Psi(\vec r)$ in the $xy$-plane. It is normalized as
$\frac{1}{\Omega} \int^\prime d^2r |\Psi(\vec r)|^2=1$ where
$\int^\prime d^2 \vec r$ integrates over one unit cell with the area
of $\Omega=2a_0^2$. The GP equation reads \bea \Big\{-\frac{\hbar^2
\vec \nabla^2}{2M}  +V(\vec r) + g \rho_0 | \Psi(\vec r)|^2\Big\}
\Psi(\vec r) =E\Psi(\vec r), \label{eq:gp} \eea where $\rho_0=N_0/V$
is the average 3D density with $N_0$ the total boson number in the
condensate and $V$ is the 3D volume of the system; $g$ is the
$s$-wave scattering interaction parameter. In the calculations
below, various values of interaction parameters $g\rho_0$ are used
from $0$ up to $E_r$.

\begin{figure}[htb]
\includegraphics[width=0.97\linewidth]{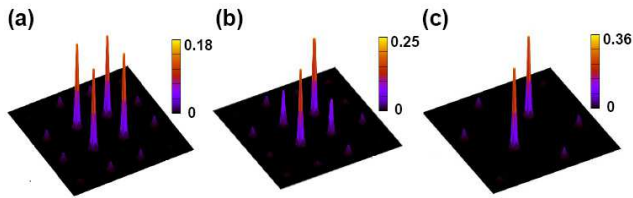}
\caption{Density distribution in the time-of-flight spectrum for
(a)complex condensate in the symmetric case ($\alpha=36.0^\circ$)
(b)complex condensate in the asymmetric case ($\alpha=35.5^\circ$).
(c)real condensate ($\alpha=34.5^\circ$). Other parameters values
are the same as Fig.\ref{fig:lattice} except $\alpha$.}
\label{fig:TOF}
\end{figure}

Although Eq. \ref{eq:gp} looks the same as the usual GP equation,
the marked difference is that $\Psi(\vec r)$ is not the ground state
condensate but the meta-stable one belonging to the second band. The
non-linearity of the GP equation allows mixing between different
Bloch wave states. Let us start from the symmetric lattice with
$\alpha=\alpha_0$. Eq. \ref{eq:gp} is solved self-consistently as
follows. We define the renormalized potential as $V_{eff}(\vec r)=
V(\vec r) + g\rho_0 |\Psi(\vec r)|^2$, and solve the corresponding
renormalized band structure. Then the condensate wavefunction is
optimized to minimize the total energy, which in turn determines
$V_{eff}$. The renormalized band structure is similar to the free
one, which still has two degenerate band minima at $K_{1,2}$. We
define the condensate wavefunction as \bea \Psi(\vec r)=\cos\delta
~\psi_{K_1}(\vec r)+e^{i\phi} \sin\delta ~\psi_{K_2}(\vec r).
\label{eq:cond} \eea The total energy reaches minimal at
$\delta=\frac{\pi}{4}$ and $\phi=\pm \frac{\pi}{2}$. These complex
condensate wavefunctions only have nodal points, while the real ones
$\psi_{K_{1,2}}$ have nodal lines. The complex ones are spatially
more uniform, and thus are favored by interactions.
We plot the phase and density patterns of this condensate in Fig.
\ref{fig:phase}, which exhibit a vortex-antivortex lattice
structure. The vortex and anti-vortex cores are located
alternatively at centers of $A$-sites, at which the
antiferromagnetic order of orbital angular momentum develops. For
every closest four $B$-sites, their phases wind around the central
$A$-site following the same vorticity.
This is similar
to the case of the tight-binding models \cite{liu2006,wu2009}. The
Bragg peaks in the time of flight (TOF) spectra are located at
$(m+\frac{1}{2}) \vec b_1  +n \vec b_2$ and $m \vec b_1 +
(n+\frac{1}{2}) \vec b_2$ as observed in the experiment
\cite{wirth2011}.
In particular, the four peaks of $\pm \frac{1}{2}
\vec b_{1,2} =(\pm \frac{\pi}{2a_0}, \pm \frac{\pi}{2a_0})$ are
strongest with equal intensities, as shown in Fig.\ref{fig:TOF} (a).

\begin{figure}[htb]
\includegraphics[width=0.485\linewidth]{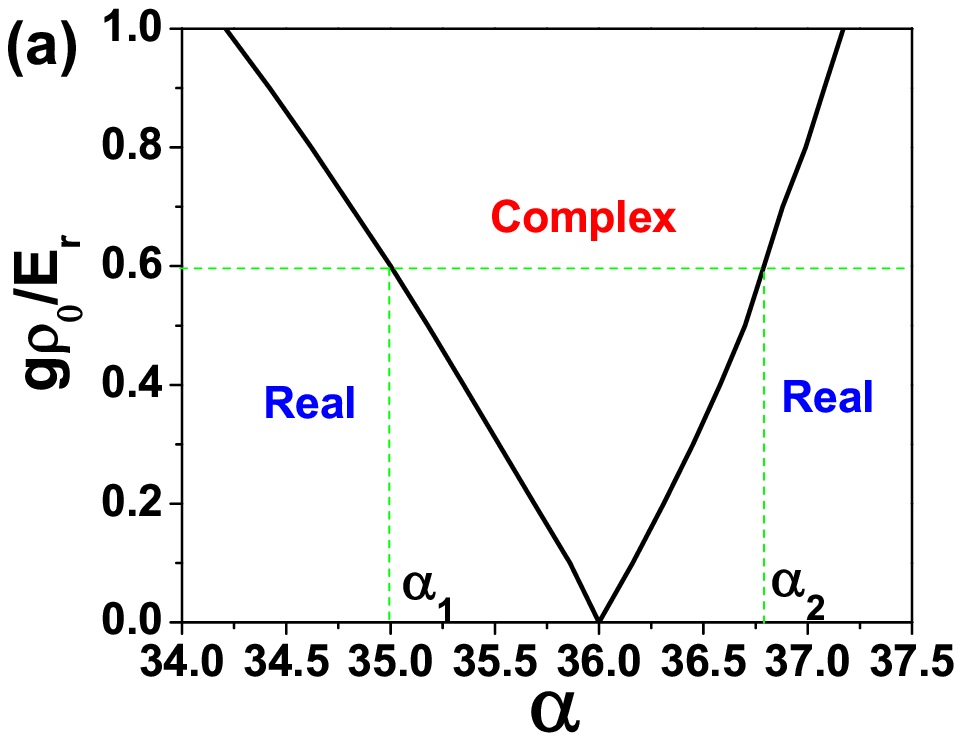}
\includegraphics[width=0.495\linewidth]{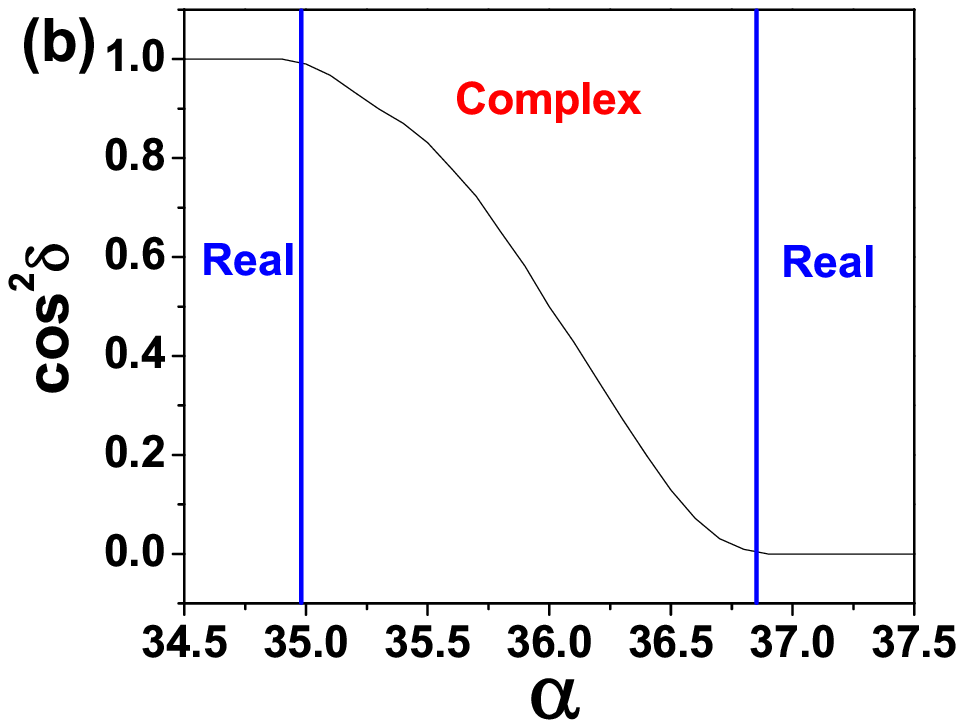}
\caption{(a)The phase diagram as $\alpha$ and the interaction
strength $g\rho_0$. Other parameters values are the same as
Fig.\ref{fig:band}. (b) The condensate fractions of $\psi_{K_1}$ in
the complex UBEC $\Psi=\cos\delta ~\psi_{K_1} \pm i \sin \delta
~\psi_{K_2}$, the parameter values are the same as
Fig.\ref{fig:lattice} except $\alpha$ and $g \rho_0=0.6 E_r$.}
\label{fig:rlvscmplx}
\end{figure}

Now we move to the asymmetric lattice whose free band structure
minimum is non-degenerate. The complex condensates are favored by
interactions, and thus should be stable at asymmetries weak enough.
Certainly, at large asymmetries, the real condensate wins due to the
gain of band energy. This picture is explicitly confirmed by the
phase diagram calculated by GP equation. As shown in Fig.
\ref{fig:rlvscmplx} (a),  for a given value of the interaction
strength $g\rho_0$, the complex condensate in the form of Eq.
\ref{eq:cond} is stable in a finite parameter range from $\alpha_1$
to $\alpha_2$, beyond this regime the condensate changes to the real
one, and  the TOF spectra of such a real condensate only contain
peaks of $(m+\frac{1}{2})\vec b_1$ or $(m+\frac{1}{2}) \vec b_2$, as
shown in Fig.\ref{fig:TOF} (c)

In the complex condensate, the relative phase $\phi$ between
$\psi_{K_{1,2}}$ is always $\pm \frac{\pi}{2}$, {\it i.e.}, $\Psi$
and $\Psi^*$ are degenerate as TR partners; $\delta$ is asymmetry
dependent. The spatial asymmetry of $|\Psi(\vec r)|^2$ depends on
that of the bare potential $V$. However, $V_{eff}$, a combination of
$V$ and $|\Psi|^2$, becomes symmetric. Without loss of generality,
$\Psi(\vec r)$ is expanded in terms of two orthonormal real
wavefunctions $\psi_{1,2}(\vec r)$ in the same way as in Eq.
\ref{eq:cond} by replacing $\psi_{K_{1,2}}$ with $\psi_{1,2}$.
Apparently, both $\Psi(\vec r)$ and $\Psi^*(\vec r)$ satisfy Eq.
\ref{eq:gp}, and yield the same $V_{eff}$. The corresponding
renormalized single particle Hamiltonian, $-\hbar^2 \nabla^2 /(2M)
+V_{eff}$, has degenerate band minima $\psi_{1,2}$. However, please
note that the superposition principle does not apply to the
non-linear GP equation: $\psi_{1,2}$ are {\it not} solutions to Eq.
\ref{eq:gp}. $|\Psi(\vec r)|^2$ is also asymmetric depending on the
asymmetry of the bare  potential $V$.
 The TOF spectra still exhibit four dominant peaks
at $\pm (\frac{\pi}{2a_0}, \frac{\pi}{2a_0})$ and
$\pm(-\frac{\pi}{2a_0},\frac{\pi}{2a_0})$, as shown in
Fig.\ref{fig:TOF} (b). The relative intensities of these two pairs
of peaks depend on the lattice asymmetry, which can be reflected by
the condensation fractions $\psi_{K_1}$ in the complex condensate,
as plotted in Fig. \ref{fig:rlvscmplx} (b). An observation of the
asymmetric peaks at $\pm \frac{1}{2} \vec b_{1,2}$ at
$\alpha_1<\alpha<\alpha_2$ would provide a supporting evidence for
the complex condensates. The TOF spectra lack phase information,
thus the observation of the symmetric peaks $\pm \frac{1}{2} \vec
b_{1,2}$ at $\alpha_0$ \cite{wirth2011} could be interpreted as the
phase separation of real condensates of $\psi_{K_{1,2}}$, or an
incoherent mixing between them. However, in these scenarios, the
lattice asymmetry lifts the degeneracy and only leads to one pair of
peaks. Even two condensates could coexist forming domains, their
nature is of hysteresis. The condensate fraction of  $\psi_{K_1}$ in
the complex condensate should not follow that plotted in Fig.
\ref{fig:rlvscmplx}(b).

\begin{figure}[htb]
\includegraphics[width=0.97\linewidth]{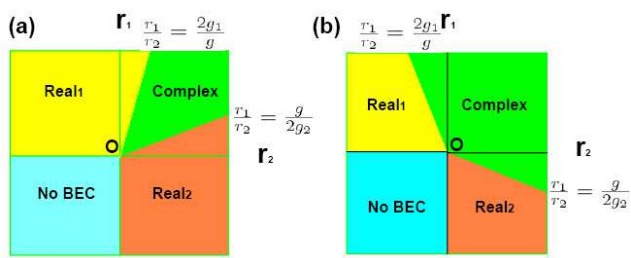}
\caption{Phase diagram as a function of $r_1, r_2$ predicted by
Eq.\ref{free2} for (a) $g>0$; (b) $g<0$.} \label{fig:phaseD}
\end{figure}

For a better understanding of phase transitions between real and
complex UBECs, we construct a Ginzburg-Landau (GL) free energy as:
\begin{eqnarray}
F&=&-r_1|\Psi_{K_1}|^2-r_2|\Psi_{K_2}|^2+g_1|\Psi_{K_1}|^4+g_2|\Psi_{K_2}|^4
\nn \\
&+&g_3|\Psi_{K_1}|^2|\Psi_{K_2}|^2+g_4(\Psi_{K_1}^*\Psi_{K_1}^*\Psi_{K_2}
\Psi_{K_2}+h.c), ~~~\label{free}
\end{eqnarray}
where $\Psi_{K_{1,2}}=\psi_{K_{1,2}}e^{i\theta_{1,2}}$ describe the
condensate order parameters at $K_{1,2}$; $\theta_{1,2}$ are the
phases of the condensates of $\Psi_{K_{1,2}}$ and $\psi_{1,2}$ are
real as explained before. Although $\Psi_{K_{1,2}}$ do not couple at
the quadratic level due to the requirement of translational
symmetry, they do couple at the quartic level as in the $g_4$ term
because $\pm 2 (\vec K_1-\vec K_2)$ equals  reciprocal lattice
vectors. $g_4$ is positive for repulsive interactions which favors
the relative phase difference $\theta_1-\theta_2=\pm \frac{\pi}{2}$,
thus the free energy in Eq.(\ref{free}) can be reduced to:
\begin{equation}
F=-r_1\psi_{K_1}^2-r_2\psi_{K_2}^2+g_1\psi_{K_1}^4+g_2\psi_{K_2}^4+g\psi_{K_1}^2\psi_{K_2}^2,\label{free2}
\end{equation}
in which $g=g_3-2g_4$. We define $G=4g_1g_2-g^2$ and $g_1,g_2,G>0$
as required by the thermodynamic stability condition. In the
superfluid regime, the complex UBEC is characterized by the non-zero
values of both $\Psi_{K_{1,2}}$, while the real BECs correspond
to one of these values being zero. Without loss of generality, we
fix $g_1,g_2,g$ and plot the phase diagram of the superfluid regime
as a function of $r_1,r_2$. As shown in Fig.\ref{fig:phaseD}, for
$g>0$, the complex BECs occur when
$\frac{g}{2g_2}<\frac{r_1}{r_2}<\frac{2g_1}g$, where both $r_1,r_2$
are positive. It is interesting to notice that for $g<0$, the
complex BEC can exist even one channel is off-critical ($r_1<0$ or
$r_2<0$), which means that in this case, the complex BEC is purely
induced by interaction.

As interaction increases, and the system is brought into the Mott
insulating regime. Nevertheless, at least in the weakly insulating
regime, the suppress of the superfluidity ordering is due to phase
fluctuations,  and the magnitudes of $|\Psi_{K_{1,2}}|$ remain
nonzero. Though $\theta_1$ and $\theta_2$ are disordered such that
$\avg{\Psi_{K_{1,2}}}=0$, their relative phase
$\theta_1-\theta_2=\pm \frac{\pi}{2}$. This indicates a TR breaking
order with a bilinear form of $\Psi_{K_{1,2}}$ as $L=i\langle
\Psi_{K_1}^*\Psi_{K_2}-\Psi_{K_2}^*\Psi_{K_1}\rangle$ in the Mott
insulating state. Its physical meaning here remains the staggered
circulating currents, i.e., this exotic Mott insulating states
preserve the antiferromagnetic OAM order of the complex BECs but not
the global phase coherence.


In summary, we have studied the UBECs observed in high orbitals
bands in Ref.[\onlinecite{wirth2011}]. The unconventional condensate
wavefunctions can be real and TR invariant with nodal lines, or
complex breaking TR symmetry with nodal points. In both cases,
translational symmetry is broken due to the nonzero condensation
wavevectors, thus these UBECs can be considered as unconventional
supersolid states. The interplay between lattice asymmetry and
interactions drives the transition between them.

We are grateful to A. Hemmerich for carefully reading our manuscript
and insightful discussions. We also  thank A. Hemmerich, C. Morais
Smith and O. Tieleman for pointing out a mistake in the earlier
version of the draft about the order of the phase transition. Z.C.
and C.W. are supported by NSF-DMR-0804775, and the AFOSR-YIP
program.



\end{document}